\documentstyle[epsfig,subeqnarray,RM97_jprart]{aipproc}

\begin{document}

\title{Cosmic Rays and Neutrinos from Gamma Ray Bursts}
\author{J\"org P. Rachen \and P. M\'esz\'aros}
\address{Pennsylvania State University, University Park, PA 16802\\
jrachen@astro.psu.edu, pmeszaros@astro.psu.edu}
\maketitle

\begin{abstract}
We review the hypothesis that the acceleration of protons at internal shocks
in Gamma Ray Bursts (GRB) could be the origin of the ultra-high energy cosmic
rays (UHECR) observed at earth, $E_{\rm max} \gsim 10^{19}\rm\,eV$. We find
that, even though protons may be accelerated to such energies, their ejection
into the interstellar/intergalactic medium is problematic because it is
likely to be accompanied by considerable adiabatic losses in the expanding
shell. The problem is circumvented by neutrons produced in photohadronic
interactions, which are not magnetically bound and thus effectively ejected
in the moment of their production. They can be both produced in sufficient
number and be able to leave the emission region if the optical depth of the
emission region to photohadronic interactions is of order 1. We show that
this requirement can be fulfilled under the same conditions which allow
acceleration of protons to the highest energies. The production of neutrinos
in this process correlates the fluxes of cosmic rays and neutrinos, and makes
the hypothesis of UHECR origin in GRBs testable.
\end{abstract}

\section*{Acceleration of Protons in GRBs}

The hypothesis that Gamma Ray Bursts (GRB) might be responsible for the
origin of ultra high energy cosmic rays (UHECR) has been proposed by Waxman
\cite{Rachen:Wax95}, who assumed cosmic ray acceleration by momentum
diffusion (2nd order Fermi acceleration), or at internal shock waves (1st
order Fermi acceleration) within the expanding wind. We investigate here in
some more detail the problems and consequences of this idea, in particular
under reference to the internal shock scenario \cite{Rachen:RM94}, where the
observed variability in GRBs is explained by the presence of shocks in an
unsteady outflow which occur when expanding subshells of different velocity
catch up and merge.

The energy of protons accelerated at internal shocks in GRBs is essentially
constrained by two conditions: (1) confinement in the acceleration region,
$r'_{\rm g} = E'_p/e B' < R' \sim c T_{\rm var} \Gamma$; (2) balance of
acceleration and synchrotron cooling, $t'_{\rm acc} \sim 2\pi r'_{\rm g}/c < t'_{\rm
syn}$. Here, $T_{\rm var}$ is the observed variability time scale, and
$\Gamma$ is the bulk Lorentz factor of the wind; primed quantities refer to
the comoving frame. It can be shown that energy losses due to photohadronic
interactions and other cooling processes are less relevant
\cite{Rachen:RM97}.  Expressing the magnetic field relative to its
equipartition value with the radiation, $U'_B = \xi_{B\gamma} U'_\gamma$, the
maximum proton energy has to satisfy the conditions
\begin{subeqnarray}\label{E:Rachen:Epmax}
\slabel{E:Rachen:Epmax:gyr} 
	E_{p,20} &\lsim& 3 (L_{51}\xi_{B\gamma})^{1/2} \Gamma_2^{-1} \\ 
\slabel{E:Rachen:Epmax:sync} 
	E_{p,20} &\lsim& 
		{\TS\frac13} (L_{51}\xi_{B\gamma})^{-1/4} T_{-1}^{1/2} \Gamma_2^{5/2} 
\end{subeqnarray}
Quantities are expressed in canonical units: $E_{p,20} = E_p/10^{20}\eV$,
$L_{51} = L/10^{51} \erg \scnd^{-1}$, $T_{-1} = T/0.1 \scnd$, and $\Gamma_2 =
\Gamma/100$.  These equations can be rewritten to express the minimum
requirements on the physical conditions in the GRB wind in order to produce
the highest energy cosmic rays:
\begin{subeqnarray}\label{cond}
\slabel{E:Rachen:gamma} 
	\Gamma_2 &\gsim& E_{p,20}^{3/4} T_{-1}^{-1/4} \\
\slabel{E:Rachen:zeta}  
	\xi_{B\gamma}  &\gsim& 0.2\,E_{p,20}^{7/2} T_{-1}^{-1/2} L_{51}^{-1}
\end{subeqnarray}
Since the highest energy cosmic rays are observed with energies up to
$3\mal10^{20}\eV$ \cite{Rachen:UHECR}, GRB winds must have bulk Lorentz
factors $\Gamma \gsim 100$ and require $U'_B\gsim U'_\gamma$.  The first
condition is remarkably close to the canonical value assumed for GRBs in the
internal shock scenario \cite{Rachen:RM94}. The latter condition is
reasonable if the cosmic ray energy density dominates about that of
relativistic electrons, $U'_p\gsim U'_e \sim U'_\gamma$, because the magnetic
field could still be in equipartition with the total energy density in
relativistic particles, $U'_B \sim U'_p + U'_e$.
 
\section*{The ejection problem for UHE protons}

For cosmic ray acceleration at internal shocks, the acceleration site is
placed within a relativistically expanding wind. The acceleration takes place
typically at radii $r_i \sim 10^{14}\cm$, while the expansion continues until
the wind hits the decelerated material behind the external shock at radii
$r_e \sim 10^{16}\cm$. In a pointing dominated flux, the magnetic field
decreases as $B'\propto r^{-1}$ and is largely transversal, because the
longitudinal component decays with $r^{-2}$. Magnetic reconnection can,
however, maintain isotropy of the magnetic field, leading to $B'\propto
r^{-2}$ \cite{Rachen:Thom94}, consistent with the generic approach of a
matter dominated flow where some mechanism keeps the magnetic field in
equipartition with the thermal gas. In a decreasing magnetic field, charged
particles suffer adiabatic energy loss due to the constancy of the adiabatic
invariant $B' {r'}_{\rm g}^2$, which leads to an energy evolution of the
nonthermal particle component with $E'\propto {B'}^{1/2}$. The condition for
adiabaticity is that the time scale of particle gyration is much shorter than
the expansion time scale, $r'_{\rm g}/c \ll r/c\Gamma$, which is equivalent
to the confinement condition during acceleration for protons sufficiently
below the maximum energy defined by $r'_{\rm g}\sim R' = r/\Gamma$. During
expansion, $r'_{\rm g}/R' \propto r^{-1/2}$ if $B\propto r^{-1}$, and
confinement during acceleration implies that even the most energetic protons
remain confined in the expanding shell and cool adiabatically. For $B'\propto
r^{-2}$, $r'_{\rm g}/R'$ remains constant and adiabaticity applies only for
protons with $E'_p \ll E'_{p,\rm max}$, but some cooling should be also
expected for $E'_p\sim E'_{p,\rm max}$ (this requires further calculations).

When the material hits the outer shell at $r=r_e$, its bulk Lorentz factor,
$\Gamma$, drops as a power law to values close to unity
\cite{Rachen:MRees97}. In this deceleration phase, the magnetic field
confinement may break up and the energetic particles can be released. Their
energy in the comoving field is then $E'_{\rm ej} \lsim (r_i/r_e)^{\alpha/2}
E'_{\rm acc}$, if $B'\propto r^{-\alpha}$, and $\Gamma_{\rm ej}\ll \Gamma_{\rm
acc}$ would additionally reduce the energy in the observers frame. Hence, we
expect a reduction of the energy of most protons {\em at ejection} by some
orders of magnitude compared to their observer frame energy immediately after
acceleration.  This would essentially rule out a dominant contribution of
GRBs to the UHECR spectrum above $10^{19}\eV$.

\section*{Neutron and Neutrino Production}

The problem of adiabatic losses can be circumvented if the ejection of
neutral particles is considered, because they are not coupled to the magnetic
field. The obvious candidates are here neutrons, which are produced by
protons in charged current photomeson-production, e.g.\ $p\gamma \to
n\pi^+$. This is the same process which is also responsible for the
production of neutrinos as a result of the pion decay, $\pi^+ \to
\mu^+\bar\nu_\mu$, $\mu^+ \to e^+\nu_\mu\bar\nu_e$. Neutrinos are produced
with an energy $E_\nu \lsim 0.05 E_n$; the neutrino energy can be
considerably below this limit, if energy losses of pions and muons are
relevant, which is the case in GRBs \cite{Rachen:RM97}. The neutrino flux
produced by this mechanism was recently proposed to reach observable levels
above $100\TeV$ \cite{Rachen:WB97}. One can show that the conditions for the
acceleration of protons to $\gsim 10^{20}\eV$ implies that neutrinos up to
$10^{16}\eV$ must be produced \cite{Rachen:RM97}.

The neutrons are left with about $80\%$ of the proton energy and carry
therefore cosmic ray energy efficiently. The production spectrum of neutrons
depends on the both the proton spectrum and the spectrum of background
photons. The relevant photon energies for the reaction are $\eps' \gsim
150\MeV {\gamma'}_p^{-1}$; for $\gamma'_p \gsim 10^5$ this is below the break
energy of GRB spectra, so that the integrated number of photons above the
reaction threshold rises only slowly with Lorentz factor. The neutron
spectrum would than be expected to follow the proton spectrum, which is
canonically assumed as $N'_p \propto {E'}_p^{-2}$; the same is the case for
the accompanying neutrino spectrum (for details see Ref.~\cite{Rachen:RM97}).
\begin{figure}
\centerline{\epsfysize=9cm\epsfbox[0 0 500 500]{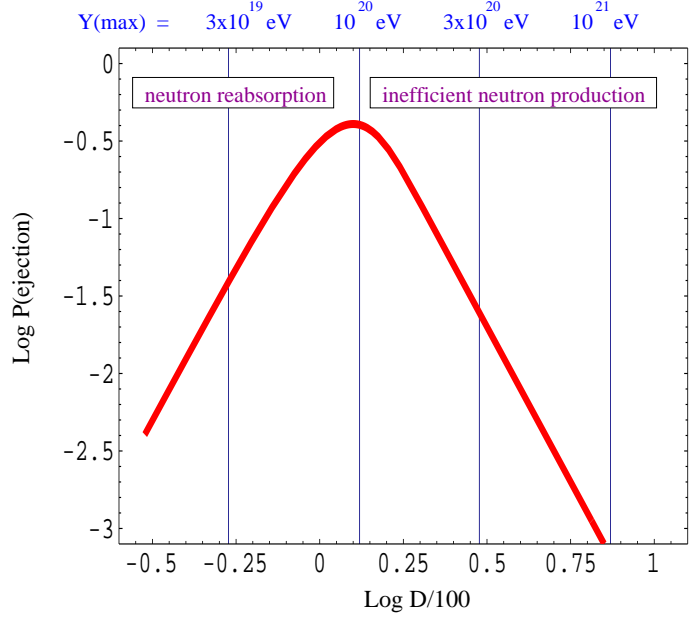}}
\caption[]{\label{F:Rachen} The ejection probability of neutrons from GRBs
vs. $D = \Gamma T_{-1}^{1/4} L_{51}^{-1/4}$, and the correlated maximum
proton energy scaled with luminosity, $Y_{\rm max} = E_{p,\rm max}
L_{51}^{-1/3}$.}
\end{figure}

In order to escape from the GRB shell, neutrons must fulfill two conditions:
Their decay time in the lab frame, $\tau_n\gamma_n$, must be considerably
larger than the total time of the burst, and the probability of reabsorption
by a $n\gamma\to p\pi^-$ reaction must be small. It is easy to see that the
first condition is satisfied in GRBs, since $c\tau_n\gamma_n > r_e$ for
$\gamma_n > 10^3$; for $\gamma_n \gsim 10^{10}$, it may even leave any
possible stronger magnetized environment of the GRB before undergoing
$\beta$-decay. Less trivial is the second condition: The probability of a
neutron to leave the GRB shell, which has a thickness $R' = r/\Gamma$, is
\begin{equation}\label{E:Rachen:Pesc}
P_{\rm esc} \sim \frac{c}{R'} \int_0^{R'/c} \exp\left(-\frac{t'}{t'_{p\gamma\to
n}}\right) \;dt' 
\end{equation}
where the integral covers the range of distances of the point of the
production of the neutron to the border of the shell. This probability is
directly related to the probability of the neutron to be produced, $P_{\rm
prod} \approx 1 - \exp(-t'_{\rm ad}/t'_{p\gamma\to n})$, where $t_{\rm ad}$
is the time scale for adiabatic cooling of the protons. In the
relativistically expanding GRB shell we have $t'_{\rm ad} \approx
R'/c$. Consequently we can write the probability $P_{\rm ej}$ for a UHECR
proton to be ejected from the GRB as a neutron as
\begin{equation}\label{E:Rachen:Pej}
P_{\rm ej} \sim \frac{c t'_{p\gamma\to n}}{R'}
\left[1-\exp\left(-\frac{R'}{ct'_{p\gamma\to n}}\right)\right]^2\quad.
\end{equation}
The characteristic ratio involved in this expression can be expressed by
canonical GRB parameters, $R'/ct'_{p\gamma\to n} \sim \frac12 \Gamma_2^4
T_{-1} L_{51}^{-1}$.  The probability $P_{\rm ej}$ as a function of $D \equiv
\Gamma_2 T_{-1}^{1/4} L_{51}^{-1/4}$ is shown in Fig.\,\ref{F:Rachen},
together with the lower limits on $D$ to produce UHECR protons. We see that
the same conditions which make GRBs perfect proton accelerators, makes them
also to almost perfect ``neutron bombs'' with ejection efficiencies of order
$1{-}30\%$.
  
\section*{Consequences}

Under the assumption that GRBs produce protons with an energy density
comparable to the radiation, and that each proton produces about 1 pion
during its lifetime, i.e.\ transferring about $ 20\%$ of the cosmic ray energy
to neutrinos, Waxman and Bahcall \cite{Rachen:WB97} have shown that GRBs can
produce a diffuse background flux of neutrinos above $100\TeV$, which would
lead to about 10 to 100 GRB correlated events per year in a $\km^3$
underground neutrino detector. This flux should be easily detectable above
the background due to the possibility of a correlation in both direction and
time with the GRB. The same conditions would suffice to contribute a large
fraction of the observed UHECR flux \cite{Rachen:Wax95}, but the connection of
cosmic ray and neutrino ejection efficiencies depends in the GRB parameters.

Here, we have argued that energetic protons cannot be emitted directly from a
GRB without losing most of their energy in adiabatic expansion, but that
neutrons produced in charged current photohadronic interactions can escape
the GRB and contribute to the cosmic ray proton spectrum after $\beta$-decay,
provided that every proton produces on average one pion or less.  This
``one-pion-requirement'' constrains the physical parameters of GRBs, but
allows the acceleration of ${\sim}\,10^{20}\eV$ protons, however, with a
strongly decreasing ejection efficiency for larger energies. One conclusion
might be that only the most luminous GRBs can produce cosmic rays of the
highest energies, $E\gsim 3\mal 10^{20}\eV$. The neutrino flux is then
one-to-one correlated to the emitted cosmic ray flux, thus VHE neutrino
observations could test the hypothesis of UHECR origin from GRBs;
non-observation of a GRB correlated neutrino flux at the level predicted by
Waxman and Bahcall \cite{Rachen:WB97} would rule out this hypothesis for
standard assumptions of UHECR propagation.

\section*{Acknowledgements}

This work was supported in part by the NASA under grant NASA5-2857.

\end{document}